\documentclass[prb,twocolumn,showpacs,preprintnumbers,amsmath,amssymb]{revtex4-1}

\usepackage{graphicx}
\usepackage{dcolumn}
\usepackage{bm}
\usepackage{color}
\usepackage{braket}
\begin{document}

\title{Flat bands and gaps in twisted double bilayer graphene}

\author{F. J. Culchac}

\address{Instituto de F\'isica, Universidade Federal do Rio de Janeiro, Caixa Postal 68528, Rio de Janeiro, RJ 21941-972, Brazil}

\author{Rodrigo B. Capaz}

\address{Instituto de F\'isica, Universidade Federal do Rio de Janeiro, Caixa Postal 68528, Rio de Janeiro, RJ 21941-972, Brazil}

\author{Leonor Chico}
\email{leonor.chico@icmm.csic.es}
\address{Materials Science Factory, Instituto de Ciencia de Materiales de Madrid, Consejo Superior de Investigaciones Cient\'{\i}ficas, C/ Sor Juana In\'es de la Cruz 3, 28049 Madrid, Spain }
\author{E. Su\'arez Morell}
\email{eric.suarez@usm.cl}
\address{Departamento de F\'isica, Universidad T\'ecnica Federico Santa Mar\'ia, Casilla 110-V, Valpara\'iso, Chile}

\begin{abstract}
We present electronic structure calculations of twisted double bilayer graphene (TDBG): A tetralayer graphene structure composed of two AB-stacked graphene bilayers with a relative rotation angle between them. Using first-principles calculations, we find that TDBG is semiconducting with a band gap that depends on the twist angle, that can be tuned by an external electric field. The gap is consistent with TDBG symmetry and its magnitude is related to surface effects, driving electron transfer from outer to inner layers. The surface effect competes with an energy upshift of localized states at inner layers, giving rise to the peculiar angle dependence of the band gap, which reduces at low angles. 
For these low twist angles, the TDBG develops flat bands, in which electrons in the inner layers are localized at the AA regions, as in twisted bilayer graphene.
\end{abstract}

\maketitle

\section{Introduction}
The stacking of a few layers in two-dimensional (2D) materials expands the possibilities to tailor their properties.  Besides ordered arrangements, which slightly alter the original monolayer symmetry, a relative rotation between layers can change the behavior of the system dramatically. Therefore, in 2D materials the relative twist angle between layers has recently become an important degree of freedom that can be exploited to obtain novel and undeniably fascinating phenomena and a new field dubbed "twistronics" is emerging \cite{Carr2017,RibeiroPalau2018}.

As a paradigmatic example, the interest in twisted bilayer graphene (TBG), which was initially motivated by 
its intriguing properties (namely, the independent-layer behavior and the reduction of the Fermi velocity observed for small angles, i.e., below $10^{\rm o}$ \cite{LopesDosSantos2007,Li2009}), has recently been boosted by the discovery of superconductivity \cite{Cao2018} and Mott insulator behavior \cite{Cao2018a} for angles close to $1^{\rm o}$. A long range moir\'e potential confines the electronic states at the Fermi level in AA-stacked regions, thus suppressing their group velocity, which drops to zero at certain magic angles   \cite{Morell2010,TramblyDeLaissardiere2010,Bistritzer2011}. These two facts collude to generate strong electron-electron interactions leading to many-body effects. In addition, since the value of the magic angle depends on the strength of the interlayer coupling, uniaxial pressure increases the coupling between layers thus augmenting the values of the magic angles and providing another knob for tuning many-body effects \cite{Carr2018,Yankowitz2018}. Such findings have attracted to the field a vast number of researchers, in the quest to explore the unconventional superconducting and correlated behavior and to explain such novel physical phenomena \cite{Gonzalez2018,Koshino2018,Yuan2018,Tarnopolsky2018}. 

In fact, the localization of electrons in certain moir\'e regions is not exclusive of graphene bilayers: In other 2D van der Waals materials, layer rotation also leads to flat bands or interesting new properties \cite{Ospina2016,Zhang2018,Wu2018,Nayak2017,Alexeev2019,Tran2019,Wu2019,Alexeev2019}. 
Other graphene-stacked systems, with more layers, can also be envisaged. Trilayer graphene, combining an AB bilayer on top of a twisted graphene sheet, has been shown to have linear and parabolic bands close to the neutrality point \cite{SuarezMorell2013}.  Optical properties dependent on the rotation angle have also been found in twisted trilayers and other twisted multilayer graphene structures; predictions that the value of the magic angles should increase with the number of layers have been made \cite{Correa2014,Capaz2018}. 

In this article we study the twisted double bilayer graphene (TDBG), composed of two AB-stacked bilayers of graphene with a rotation between them. This system has been recently the focus of a few experimental and theoretical works \cite{Choi2019,Liu2019,Cao2019,Koshino2019,Haddadi2019}. 
We find by density functional theory (DFT) calculations that TDBG has a 40 meV gap at the K point of the Brillouin zone for large angles, which decreases for decreasing angles. The origin of this dependence is a combination of surface effects and an angle-dependent relative shift of Dirac points of inner and outer layers.
More specifically, the semiempirical tight binding (TB) methods that were successfully employed on twisted bilayer structures fail to reproduce the DFT band gap for large angles in the TDBG.
In order to agree with DFT results, 
a constant, layer-dependent onsite term has to be added to the TB Hamiltonian to take into account the different chemical environments of inner and outer layers (surface effect) \cite{Haddadi2019}. We find that
including these onsite terms is important to reproduce the charge transfer between layers observed in DFT calculations.    
With this modification, the angle-dependence of the gap is captured naturally by the TB calculations. 
The combination of a shift of the Dirac point of the inner twisted bilayer with respect to the outer layers and the electrostatic onsite potential are the two competing  mechanisms behind behavior of the gap vs. rotation angle. 
Our calculations also show that TDBG is a direct semiconductor down to 2$^{\rm o}$ when the gap becomes indirect. 
 We also confirm that there are magic angles in TDBG which are slightly larger than those of TBG. For angles around 1.2$^{\rm o}$, the system becomes metallic and flat bands develop near the Fermi level, even though the gap at the K point persists. For angles below the magic angle, the gap can be reopened and controlled by the application of an external electric field. 

   \begin{figure}[thpb]
      \centering
\includegraphics[width=\columnwidth]{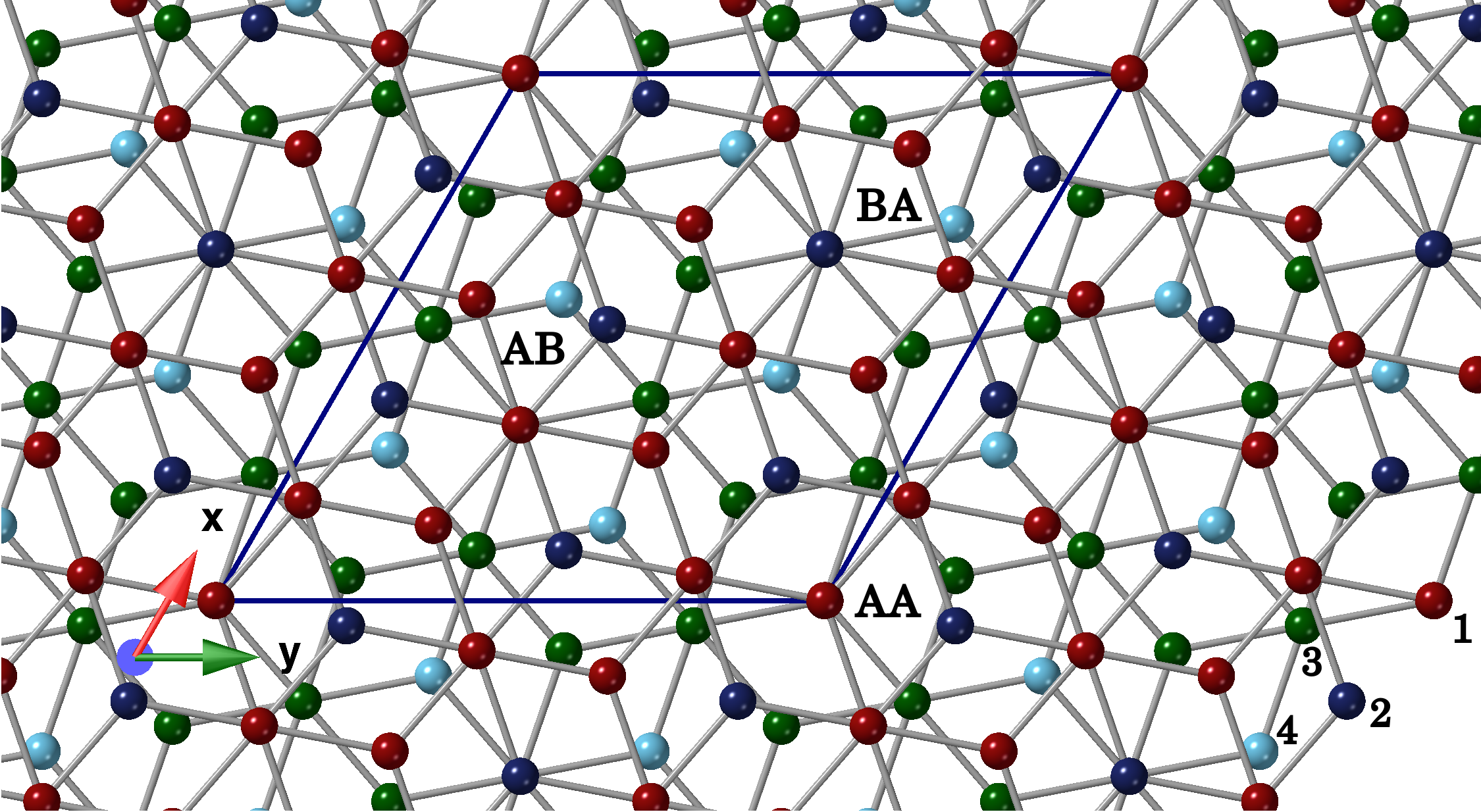}
\caption{\label{fig:UC}
Unit cell of the double AB twisted four-layer graphene (ABBA) for a relative rotation angle of 21.8$^{\rm o}$. Each layer has a different color and the numbers label the layer. The stacking between 1-2 and 3-4 is Bernal, AB and BA, and the relative rotation angle is between layers 2-3. AA, AB and BA are regions of interest in the moir\'e pattern.
} \end{figure}

\section{Theory}
\subsection{Unit cell}

Fig. \ref{fig:UC} depicts the geometry of a particular TDBG with twist angle of 21.8$^{\rm o}$. Our TDBG construction proceeds by initially stacking four graphene layers in the ABBA arrangement,  in such a way that the twisted inner layers have direct stacking at zero twist angle. 
Subsequently, the top two layers are jointly rotated around an axis going through a pair of AA atoms belonging to the two outer layers \cite{SuarezMorell2011} to build a commensurate unit cell. Therefore, in the final structure, the stackings between layers 1-2 and 3-4 are AB and BA, while layers 2-3 are rotated by an angle between 0$^{\rm o}$ and 30$^{\rm o}$ so the whole structure is periodic in a larger supercell. The point group symmetry of this structure is $D_3$, like in TBG: It has an out-of-plane threefold rotation axis ($C_{3z}$) and three in-plane twofold rotation axes $C_{2}$, two of them along the blue lines delimiting the unit cell and the third one at 60$^{\rm o}$ from them. Notice that in this setup the origin of the unit cell is at the site AA, located at the lower left corner of the unit cell highlighted in blue in Fig. \ref{fig:UC}.

We have verified that the electronic properties at low energies do not change if we start from an ABAB tetralayer with Bernal stacking at zero twist angle.
Depending on the position of the rotation axis, ABAB tetralayers can have either $D_{3d}$ or  a lower $C_3$ symmetry.  Likewise, an ABBA tetralayer rotated around an axis passing through the B sites has $D_{3d}$ symmetry. 
These various possibilities correspond to translations of layers 1-2 with respect to layers 3-4. There are differences at larger energies due to the different symmetries of the stackings, which consist of crossings or anticrossings in other energy ranges. These details are discussed in
Appendix 
\ref{Symm}.

\subsection{DFT calculations}
DFT calculations were done using Quantum Espresso (QE) allowing the structure to relax \cite{Giannozzi2009}.  
The projector augmented wave (PAW) method \cite{Corso2014} and PBE exchange correlation functional \cite{PBE} were employed; van der Waals interactions were included for all structures with the DFT-D2 method \cite{Grimme}. A grid of 6$\times$6$\times$1 $k$ points was used to relax the structures and a finer grid of up to 16$\times$16$\times$1 to obtain the total energies and band structures with a convergence threshold of $10^{-8}$ eV. 

   \begin{figure}[thpb]
      \centering
\includegraphics[angle=0,width=\columnwidth]{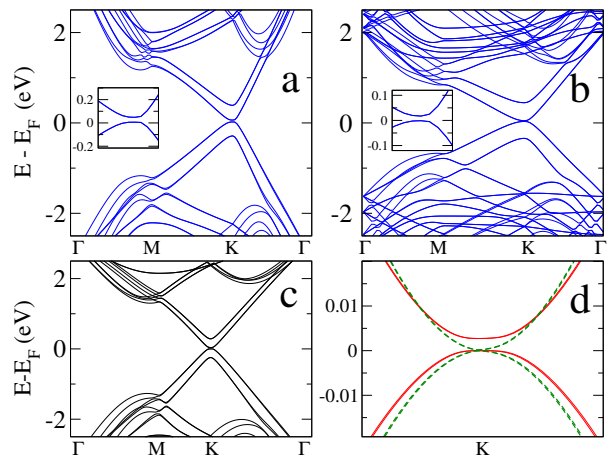}
\caption{\label{fig:dft}
 Band structures calculated with QE for two rotational angles; (a) 21.8$^{\rm o}$ and (b) 13.1$^{\rm o}$. The insets show a zoom around the K point. (c) Band structure calculated with a full TB model for the 21.8$^{\rm o}$ TDBG. (d) Comparison of the low-energy behavior of the 21.8$^{\rm o}$ TDBG calculated with two TB models, one including only nearest intralayer hopping (dashed green) and the other with more intralayer hoppings with a cutoff radius (red); for both we have set the on-site term $\Delta=0$ (see text). 
}
   \end{figure}
We have also checked that the difference in energy between the ABBA and ABAB structures is less than 0.1 meV/atom. Therefore, it is safe to say that they are both equally  stable. We have also calculated the average distance between the twisted layers after relaxation and for the 13.1$^{\rm o}$ structure it is  0.15 \AA\   larger than between the two stacked AB layers. This result is expected, as the interlayer distance is maximum for AA-stacked graphene, and in TBG it should decrease for decreasing rotation angles, reaching a minimum for AB bilayer graphene \cite{Kwanpyo2012}. 
   
\subsection{Tight-binding model} 

DFT calculations are beyond our capabilities for low angles, which have large unit cells. As in TBG, the tight-binding method is a reliable technique to obtain the electronic properties of such systems. 
All the details of the parameters can be found in the Appendix \ref{TBFull}. Additionally, a simplified model, employed to clarify the origin of the gap and its angle dependence, is given in Appendix \ref{TBSimple}. 

In order to give a good description of the band structure of TDBG, we employ a Hamiltonian with exponentially decaying hopping matrix elements between layers, either AB-stacked or twisted, 
\cite{TramblyDeLaissardiere2010,SuarezMorell2011,Nam2017}
 where only adjacent layers are coupled. 
Initially we considered only one intralayer nearest-neighbor hopping. 
This TB Hamiltonian has been used previously, giving excellent results for bilayer and trilayer graphene structures  \cite{SuarezMorell2013,SuarezMorell2015,SuarezMorell2017}. 
However, in the present work we find that the TB model with only nearest-neighbor intralayer hopping fails to reproduce the band gap found by first-principles calculations, giving an order of magnitude smaller value for large angles. 

Since the band gap opens when more than nearest-neighbors 
inside a layer are considered, 
as done by other authors, 
\cite{TramblyDeLaissardiere2010,TramblyDeLaissardiere2012,Nam2017,Capaz2018} we explore this possibility for TDBG. 
We have included more intralayer neighbors with an exponential decay and a cutoff radius, as in the interlayer case. We have tested several cutoff radii, analyzing the band gap dependence on the hopping terms.
Despite a general tendency for the band gap to increase upon inclusion of longer-range intralayer hoppings, for reasonable values of these hoppings  
the TB gap is still very small compared to that given by DFT. 

This serves as motivation to consider the inclusion of extra onsite diagonal terms with different values in inner and outer layers (see Appendix \ref{TBFull}) in the TB model \cite{Haddadi2019}. This is reasonable since outer and inner layers see different chemical environments, a surface effect that will certainly produce some amount of charge transfer between them.  
The validity of this hypothesis can assessed with the DFT calculation, as we discuss in the next Section.

\section{Results and discussion}
\label{res}

The band structures obtained using QE for TDBG (Fig. \ref{fig:dft}) resemble those of AB bilayer graphene; this indicates that a model of two weakly interacting AB bilayers is a good approximation, at least for large angles. However, as mentioned above, a closer look reveals gaps around 40 meV for  
$\theta = 21.8^{\rm o}$ (Fig. \ref{fig:dft}a) and 30 meV for $\theta = 13.1^{\rm o}$ (Fig. \ref{fig:dft}b).
Notice that the low-energy bands are degenerate except at the M-$\Gamma$ direction, like in TBG. Fig. \ref{fig:dft}c shows the band structure of a $\theta = 21.8^{\rm o}$ TDBG calculated with a TB Hamiltonian including an onsite term of  $\Delta=20$ meV, with opposite signs for inner (negative) and outer (positive) layers; they are in excellent agreement with the DFT results.

To further explore the band structure dependence on the details of the TB model, we have calculated the TB band structure for the 21.8$^{\rm o}$ TDBG with two models, in one case including only nearest intralayer hoppings and in the other case including more intralayer interactions within a 5$a_0$ cutoff radius ($a_0$ is the carbon-carbon distance), in both cases considering $\Delta=0$. In Fig. \ref{fig:dft}d we show the results at low energy around the K point. The gap is almost zero for the calculation with nearest only intralayer hopping and it increases when more intralayer neighbors are included. However, even in this case, the gap is only 6 meV, much smaller than the DFT value. This confirms that surface effects (onsite term $\Delta$) are crucial for a proper description of TDBG band structure.

    \begin{figure}[thpb]
      \centering
\includegraphics[width=\columnwidth]{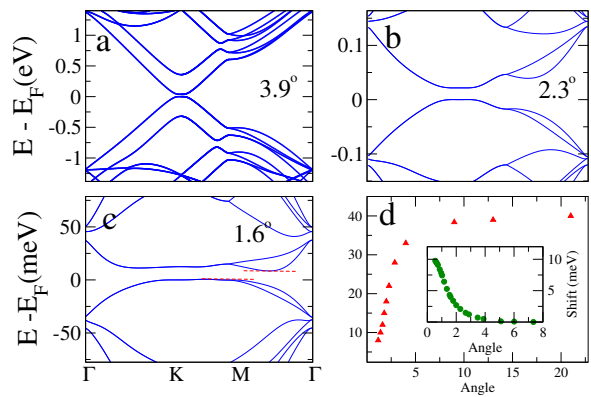}
\caption{\label{fig:TB1}
(a-c) Band structures calculated with TB for three different twist angles. (d) The value of the gap as a function of the angle. The gap is indirect below 2$^{\rm o}$. In the inset shift of the Dirac cone in twisted bilayer. The red dashes lines in (c) indicate the indirect gap.  
}
   \end{figure}
   
We calculate then the band structures for small-angle TDBGs (larger unit cells) with the full TB model including the onsite $\Delta$ terms. 
In Fig. \ref{fig:TB1} we show the bands for three different angles.
For angles below $2^{\rm o}$, due to the band dispersion in the $\Gamma$-M direction, the gap becomes indirect and the system eventually becomes metallic at $1.5^{\rm o}$. Fig. \ref{fig:TB1}(c) shows that for 1.6$^{\rm o}$ the the bands closer to the Fermi level become flat, signaling the proximity of the magic angle. We mark with dashed lines the indirect gap. Notice, however, that the two sets of flat bands never cross, i.e. for a given wavevector there is always a gap between them. In Fig.  \ref{fig:TB1}d we plot the angle dependence of the band gap at the K point (red triangles): It shows a monotonically decreasing behavior as the twist angle decreases, with a sharp drop for angles below 5$^{\rm o}$. We explain this behavior below.  


   \begin{figure}[thpb]
      \centering
\includegraphics[width=\columnwidth]{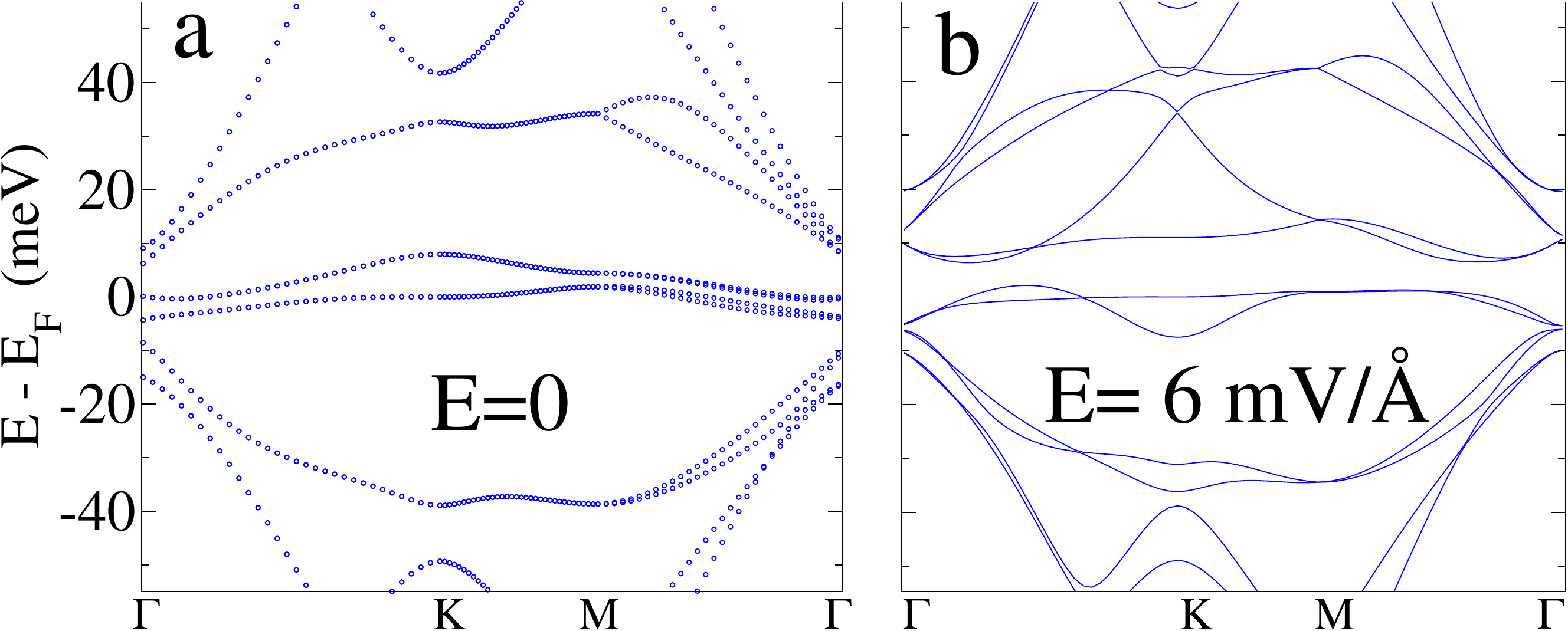}
\caption{\label{fig:magic}
Band structures calculated with the TB model for an angle of 1.16$^{\rm o}$, (a) without external electric field; (b) with a perpendicular electric field of 6 mV/\AA .   
}
   \end{figure}

At variance to what happens in TBG, our calculations reveal that the flat bands are present for a range of angles between 1.1$^{\rm o}$-1.23$^{\rm o}$. We show in Fig. \ref{fig:magic}a the band structure for a twist angle of $1.16^{\rm o}$: The system is metallic, but the two sets of nearly doubly-degenerate flat bands are separated from the valence and conduction bands.
This behavior is similar to that reported by Choi and Park \cite{Choi2019}, but different to other models for TDBG \cite{Liu2019,Cao2019}. It is also in contrast to the flat bands in TBG, which present a maximum gap at $\Gamma$ and flat bands appearing in the K-M direction \cite{Koshino2018}.

Most importantly, unlike TBG, in TDBG it is possible to open and control the size of the gap by an external electric field \cite{Choi2019,Cao2019}. As an example, Fig. \ref{fig:magic}b shows how the gap is open for the $1.16^{\rm o}$ ABBA tetralayer with an external electric field of 6 mV/\AA\  perpendicular to the layers. The four initially flat bands are split in two groups by this gap and their width enhance considerably, with the higher unoccupied bands displaying considerable mixing with the other conduction bands. 





\begin{figure}[thpb]
      \centering
\includegraphics[width=\columnwidth]{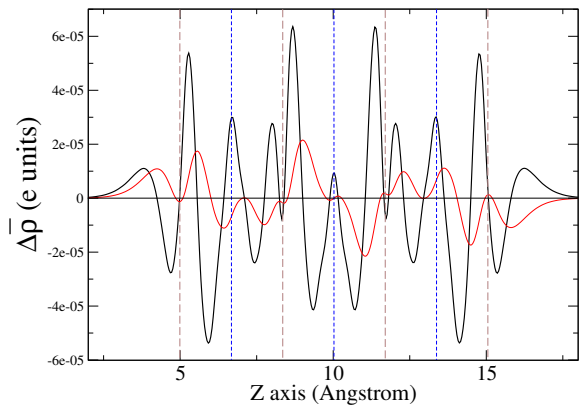}
\caption{\label{fig:EC}
Average of the charge density difference $\overline{\Delta \rho}$ along the $z$ axis of the tetralayer structure (black line) in units of the electron charge (e). The brown vertical dashed lines mark the positions of the layers. The blue dashed lines show the middle point between layers; the red curve is the cumulative integral of the charge density.     
}
   \end{figure}

As we have argued, the need for an onsite term $\Delta$ to reproduce the DFT results is a consequence of the inequivalence between inner and outer layers (surface effect), which is crucial for a quantitative description of the band gap in TDBG. Such inequivalence should also lead to an electronic charge transfer between layers. To verify this hypothesis we have estimated by DFT calculations the redistribution of the electronic density among the four layers. We first calculated the electronic density for the tetralayer structure and subtracted the electronic density of the four individual layers 
calculated independently.
The result is a spatial density of excess/depletion of electrons shown in Fig. \ref{fig:EC} . We can assign a charge to each graphene layer simply by partioning the space along planes located exactly halfway between layers.

In Fig. \ref{fig:EC} the black line shows how the average charge density difference (CDD) changes with $z$. The brown vertical dashed lines indicate the positions of the graphene layers and the blue dashed-lines the middle point between them. 
The total charge gained or lost by one layer is obtained from the integral of the CCD between the $z_l-d/2$ and $z_l+d/2$, where $z_l$ is the position of the layers and $d=0.335$ nm the distance between layers. 
Using this procedure, we estimate that the  average electronic density in each layer is $\sigma = \pm 1.8 \times 10^{-4} e / {\rm \AA}^2$ (outer layers are depleted of electrons, whereas inner layers correspondingly gain this electron density). This is consistent with the signs (+ $-$ $-$ +) of the onsite term $\Delta$
introduced in the TB model, as electrons are expected to move to regions of lower potential energy. 

The existence of a band gap in TDBG is consistent with group theory, considering a model of "weakly coupled" AB bilayers. 
An isolated AB bilayer is gapless 
due to sublattice-and-layer-exchange symmetry resulting in a $D_{3d}$ symmetry, which has at most two-dimensional irreducible representations (irreps). The group of wavevector at the K point for the AB bilayer is $D_3$, and it also contains at most two-dimensional irreps. Therefore, only doubly-degenerate states are allowed at the Fermi level. However, 
 if we bring two bilayers together to form a 4-layer TDBG structure, in the limit of zero coupling (infinite distance) between bilayers, one would have a four-fold degenerate state at the Fermi level. Thus, as soon as a small coupling is "turned on" between the two AB bilayers, level repulsion occurs and either a gap opens or two of the bands have to split off, because the $D_3$ group does not allow four-dimensional irreducible representations. Similar arguments apply for TDBG structures with $C_3$ symmetry, which also have at most two-dimensional irreps. In fact, the sublattice-and-layer-exchange symmetry is broken by any perturbation that allows to distinguish between layers, such as a potential difference \cite{Ohta2006,Oostinga2008,Pelc2015_2} or another layer(s) coupled with a twist \cite{SuarezMorell2013}. This symmetry breaking results in the opening of a gap between the AB-related bands, due to the different atomic coordination environments of the two AB layers.

Now we explain why the gap shrinks for low angles. In essence, this is related to an energy upshift due to the onset of localization of electronic states at the AA regions as the twist angle gets smaller and flat bands develop. This effect also arises in simpler TBG, but in that case it has no other consequence than a rigid displacement of the band structure. For TDBG, we can probe this effect in our TB model by artificially uncoupling the AB-stacked bilayers. This results in an inner TBG plus two outer (and isolated) layers of graphene, which serve as probes of the electronic structure. 
The bands for 
the artificially uncoupled system consist of two Dirac cones: The lower one, with higher Fermi velocity, corresponds to the two isolated graphene layers, whereas the upper one, with a renormalized Fermi velocity, corresponds to the inner TBG (see Appendix \ref{TBSimple}). 
As soon as the interlayer coupling is reintroduced in the AB-stacked bilayers, 
the original TDBG gap is recovered and, remarkably, the gap at the K point is precisely positioned at the two Dirac points of the uncoupled system. This explicitly shows that the gap in the parabolic bands is directly correlated with the upshift of the twisted bands.

\begin{figure}[thpb]
      \centering
\includegraphics[width=\columnwidth]{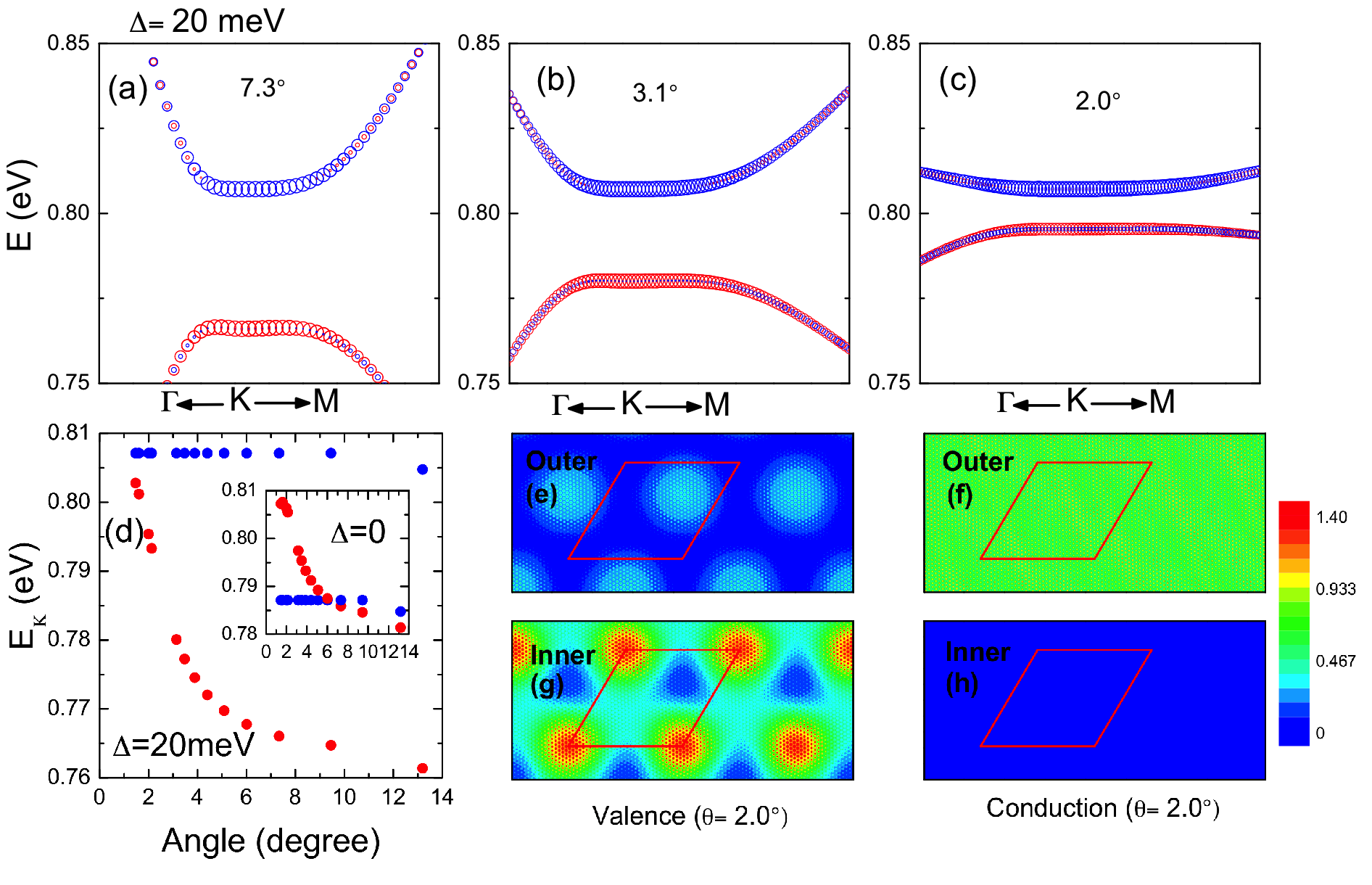}
\caption{\label{fig:ldos}
(a)-(c) Band structure near the K point for different twist angles. The sizes of red and blue dots indicate the degree of localization in the inner and outer layers, respectively.  (d) Energies of valence and conduction band states at the K point for $\Delta =$ 20 meV (main plot) and $\Delta = $0 (inset). The colors red and blue indicate the main layer-localization character of the states (inner or outer, respectively). (e)-(h) 
LDOS at the K point for $\theta = 2.0 ^{\rm o}$ in the following layer/band sequence: outer/valence, outer/conduction, inner/valence and inner/conduction. Valence states 
are 
primarily 
localized 
at AA sites for inner layers, with some mixing from AB sites at outer layers. On the other hand, conduction band states are fairly delocalized in-plane and strongly concentrated at outer layers. The color code for density is the same for all four panels.
}
   \end{figure}

We can analyze the angle-dependence of this behavior by looking at the spatial distribution of electronic states near the gap, shown in Fig. \ref{fig:ldos}. Figs. \ref{fig:ldos}(a)-(c) show a sequence of band structure plots for several angles (7.3$^{\rm o}$, 3.1$^{\rm o}$ and 2.0$^{\rm o}$) near the K point, in a very narrow energy range. The plots also carry information on the layer localization of electronic states: The sizes of red dots indicate the amount of localization in the inner layers, whereas the sizes of blue dots indicate localization in the outer layers. Notice that valence bands are primarily localized at the inner layers for most angles, but some mixing with outer layer states can be seen for the smaller angle (2.0$^{\rm o}$). More interestingly, Fig. \ref{fig:ldos}(d) shows that, as the twist angle decreases, the valence bands (inner layers) shift up as the energy of conduction bands (outer layers) remain approximately constant. This is consistent with the idea of energy shifting of Dirac points described above. In the inset of this figure, we show the corresponding behavior for $\Delta=0$. In that case, there is actually an anticrossing of bands associated to inner and outer layers, leading to a closure and subsequent reopening of the band gap at small angles, as observed by Choi et al. \cite{Choi2019}.


We now analyze the in-plane spatial localization of electronic states in TDBG. In the case of TBG, electronic states associated to angle-dependent van Hove singularities observed in TBG \cite{Li2009,TramblyDeLaissardiere2012} are revealed in scanning tunneling microscopy (STM) measurements as bright spots in the AA regions of the supercell. Figs. \ref{fig:ldos}(e)-(h) show the
local density of states (LDOS) 
in ABBA TDBG for both valence and conduction flat bands 
at the K point for $\theta=2.0^{\rm o}$.  Since the two outer layers on the one hand and the two inner layers on the other hand display a similar behavior, we depict the result for only one of each. 
We see that the valence band states display strongly localized states in AA regions of the inner layers and a smaller component of AB-localized states at the outer layers. Contrarily, the conduction band states show no in-plane localization at all, the LDOS spreads over the outer layers in a rather uniform manner. The different in-plane localization behaviors of valence and conduction bands is consistent with the different twist-angle dependence of their energies at K and they should have implications for the doping dependence of transport properties in these systems. In addition, these localization patterns call for 
a 
more precise 
approach than what was previously used \cite{Koshino2018} in the construction of a low-energy Hamiltonian based on localized Wannier functions.

We emphasize that the upshift of inner layer states as the angle diminishes arises naturally from the tight-binding model, without the need of including any angle-dependence on the TB parameters. However, in order to describe this effect in an effective low-energy model, we resort to a modified version of the model proposed recently by Koshino \cite{Koshino2019}, in which we add a $\Delta$ term plus the inner-layer shift $\epsilon_0 (\theta)$, yielding 
the following Hamiltonian for the K valley:
\begin{equation}
H=\begin{pmatrix}
H_1(k_1) & U_{AB} &  & \\ 
U^{\dagger}_{AB} & H_2(k_1) & U_{T}^{\dagger} & \\ 
 & U_{T} & H_3(k_2) &  U_{AB}\\ 
 &  &  U^{\dagger}_{AB} & H_4(k_2)
\end{pmatrix}
\end{equation}
where $k_{1,2}=R(\pm \theta/2)(k-K)$, (K- high symmetry point of the Brillouin zone) $U_{AB}$ and $U_{T}$ are the bilayer AB and moir\'e interlayer coupling respectively and the Hamiltonians for each graphene layer $H_\eta$, ($\eta=1-4$) is given by
\begin{equation*}
 H_1=   \begin{pmatrix}
\Delta & -\hbar v k_{-} \\ 
-\hbar v k_{+} & \Delta
\end{pmatrix}     ,   H_2=\begin{pmatrix}
-\Delta + \epsilon_0(\theta) & -\hbar v k_{-} \\ 
-\hbar v k_{+} & -\Delta + \epsilon_0(\theta)
\end{pmatrix} 
\end{equation*}

\begin{equation*}
 H_3=   \begin{pmatrix}
-\Delta + \epsilon_0(\theta) & -\hbar v k_{-} \\ 
-\hbar v k_{+} & -\Delta+ \epsilon_0(\theta)
\end{pmatrix}    ,   H_4=\begin{pmatrix}
\Delta  & -\hbar v k_{-} \\ 
-\hbar v k_{+} & \Delta 
\end{pmatrix} 
\end{equation*}
where $k_{\pm}=k_x \pm i k_y$, $v$ is the band velocity of graphene, $\Delta$=20 meV and $\epsilon_0(\theta)$ is an angle-dependent term accounting for the shift of the bands associated with the inner twisted layers, only arising at low angles. We show in the inset of Fig. \ref{fig:TB1}d its behavior, which mimics the gap closure at low angles. 
Note the signs of $\Delta$ in the different layers and that $\epsilon_0(\theta)$ is positive and it is only present in the inner layers. As a result, the combination of the two parameters $\Delta$ and $\epsilon_0$ reflects two competing effects: As $\epsilon_0$ increases, the gap reduces (similar to the $\Delta=0$ model). In fact, we have verified that these effects are truly competing: if in the TB calculation we reverse the signs of the $\Delta$ terms in the TDBG, the gap does not close.  

In summary, we have found using DFT and TB calculations that TDBG shows a relative large band gap of 40 meV for large twist angles. The gap is consistent with predictions from group theory and its size has an important contribution from electrostatics, as the chemical inequivalence between inner and outer layers (surface effect) drives electrons towards inner layers,  as we have verified by DFT calculations. The simple TB model fails to reproduce this effect gap, so layer-dependent on-site $\Delta$ terms have to be introduced. The gap then shrinks for smaller angles as a result of two competing effects: The surface effect plus an energy upshift of localized states at the inner twisted layers.  For low angles the gap becomes indirect 
and TDBG becomes metallic with flat bands, as in TBG; 
More importantly, an external electric field can be used to open and tune a gap between the flat bands.
Flat-band electrons and holes in TDBG are mostly localized in the AA regions of the inner twisted layers.


\section*{Acknowledgements}
ESM acknowledges financial support from FONDECYT Regular 1170921 (Chile). Powered@NLHPC: This research was partially supported by the supercomputing infrastructure of the NLHPC (ECM-02), and by Spanish MINECO, AEI and MCIU  and the European Union under Grants No. FIS2015-64654-P (MINECO/FEDER, UE) and  PGC2018-097018-B-I00 (MCIU/AEI/FEDER, UE). F.J.C. and R.B.C  are grateful to the Brazilian funding agencies CAPES, CNPq, FAPERJ and INCT - Nanomateriais de Carbono for financial support and to the High Performance Computing Center (NACAD), COPPE, UFRJ for the use of supercomputing facilities. ESM thanks Luis Brey and Eduardo Men\'endez for helpful discussions.

\appendix

\begin{center}
\textbf{\large Appendices}
\end{center}
\setcounter{figure}{0} 
\setcounter{section}{0} 
\setcounter{equation}{0}
\setcounter{page}{1}
\renewcommand{\thepage}{S\arabic{page}} 
\renewcommand{\thesection}{S\Roman{section}}   
\renewcommand{\thetable}{S\arabic{table}}  
\renewcommand{\thefigure}{S\arabic{figure}} 
\renewcommand{\theequation}{S\arabic{equation}} 

\section{Band structure and symmetries}
\label{Symm}
\begin{figure}[thpb]
      \centering
\includegraphics[width=\columnwidth]{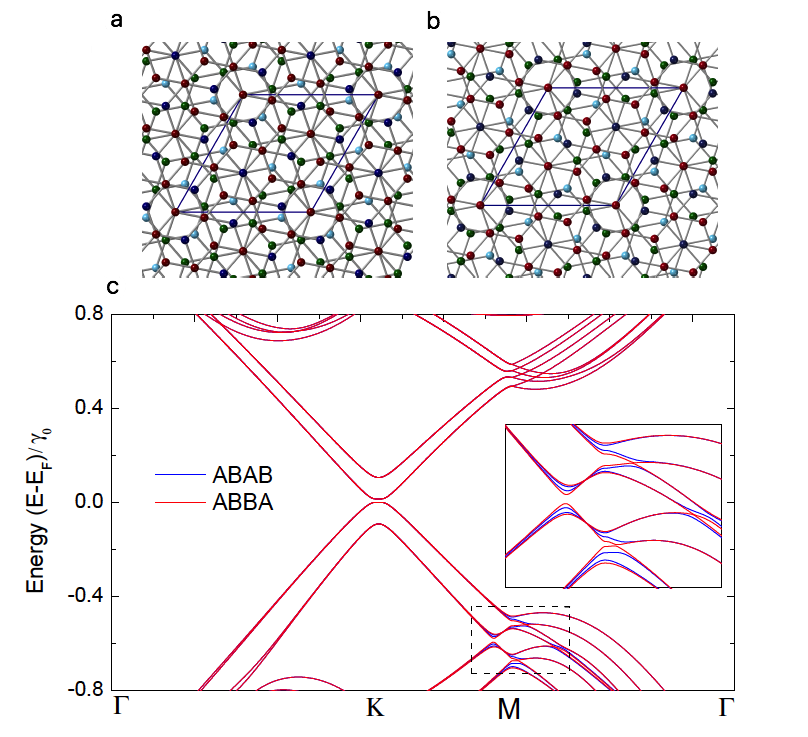}
\caption{\label{fig:A1}
Comparison of the band structures for two possible stackings (rotational angle=21.8$^{\rm o}$). (a) ABAB and (b) ABBA. The rotation axis goes through atoms AA of layers 1 and 3 in the ABAB case and through  AA  atoms of layers 1 and 4 in the ABBA case. (c) Band structures calculated with a TB model in both cases. The inset shows differences at high energies.
} 
   \end{figure}

We have considered two different geometries of the unit cell with different symmetries. If the moir\'e supercell is obtained starting from an ABBA stacking with the rotation axis going through the A sites, such that at zero twist angle the inner layers are in direct stacking, then the crystal has a $C_{3z}$ plus three $C_{2}$ symmetries (180$^{\rm o}$ rotations through an axis parallel to the layers). However, starting from an ABAB stacking with the same axis of rotation, the $C_{2}$ symmetry is not present. In order to keep the $C_{2}$ symmetry in the ABAB stacking, the rotation axis should pass through the B atoms. 
Fig. \ref{fig:A1} shows the two twisted cells starting from an ABAB (a) and ABBA (b). In Fig. \ref{fig:A1}a no $C_{2}$ symmetry can be found.

The bands of the two $21.8^{\rm o}$ structures (Fig. \ref{fig:A1}c) show some differences at the M point of the Brillouin zone at high energies. 
In the region of interest for the flat bands, i.e., at low energies, there are not perceptible differences. The crossings and anticrossings are due to the presence or absence of $C_{2y}$ symmetry.

\section{Full TB model}  
\label{TBFull}
The full TB Hamiltonian employed in the article, including the on-site terms,  is an extension of a previous model \cite{TramblyDeLaissardiere2010}

\begin{equation*}
 H = \sum_i \Delta_i \ket{i}\bra{i} - \sum_{i \neq j} t(\bold{R}_i-\bold{R}_j)  \ket{i}\bra{j}
\end{equation*}
where $\Delta_i=\pm$ 20 meV. The on-site term takes a positive or negative value depending on the layer, and we ultimately relate it to the charge transfer. 
The distance-dependent hopping parameter $t(\bold{d})$ is given by
\begin{equation*}
 -t(\bold{d})=V_{pp\pi}(d)\left[ 1- \left( \frac{\bold{d\cdot \bold{e}_z}}{d}\right)^{2} \right] + V_{pp\sigma}(d)\left(\frac{\bold{d\cdot e}_z}{d}\right)^{2}  ,
\end{equation*}

\begin{equation*}
V_{pp\pi}(d)=V_{pp\pi}^{0} exp\left(-\frac{d-a_0}{r_0}\right)  ,
\end{equation*}
\begin{equation*}
V_{pp\sigma}(d)=V_{pp\sigma}^{0} exp\left(-\frac{d-d_0}{r_0}\right)  ,
\end{equation*}
where $\bold{d}=\bold{R}_i-\bold{R}_j$ is the distance between two atoms, $e_z$ is the unit vector on the $z$ axis. $V_{pp\pi}^{0} = -2.79$  eV is the intralayer nearest-neighbor hopping in graphene, $a_0=$0.142 nm is the nearest-neighbor C-C  distance, $V_{pp\sigma}^{0}$=-0.14 $V_{pp\pi}^{0}$ is the interlayer nearest-neighbor hopping, and $d_0 =0.335$ nm is the distance between consecutive layers. The decay length parameter $r_0$ is chosen to be 0.45 nm to control the value of the second nearest-neighbor hopping. Notice that the parameter $\epsilon_0$ included in the low-energy Hamiltonian is not needed here, because the electron-hole symmetry is broken for this model.      

\section{Simplified TB model}  
\label{TBSimple}
To understand the origin of the gap we did several tests by varying the number of intra- and interlayer hopping parameters in the TB model. We noticed that if only one intralayer  hopping $\gamma_{0}$ is included, then at large angles there is almost no gap, due to the fact that this approximate Hamiltonian is more symmetric. However, for low angles the gap is open within this model. The origin of the gap opening is related to a shift in energy of the bands of the twisted inner layers.

   \begin{figure}[thpb]
      \centering
\includegraphics[width=\columnwidth]{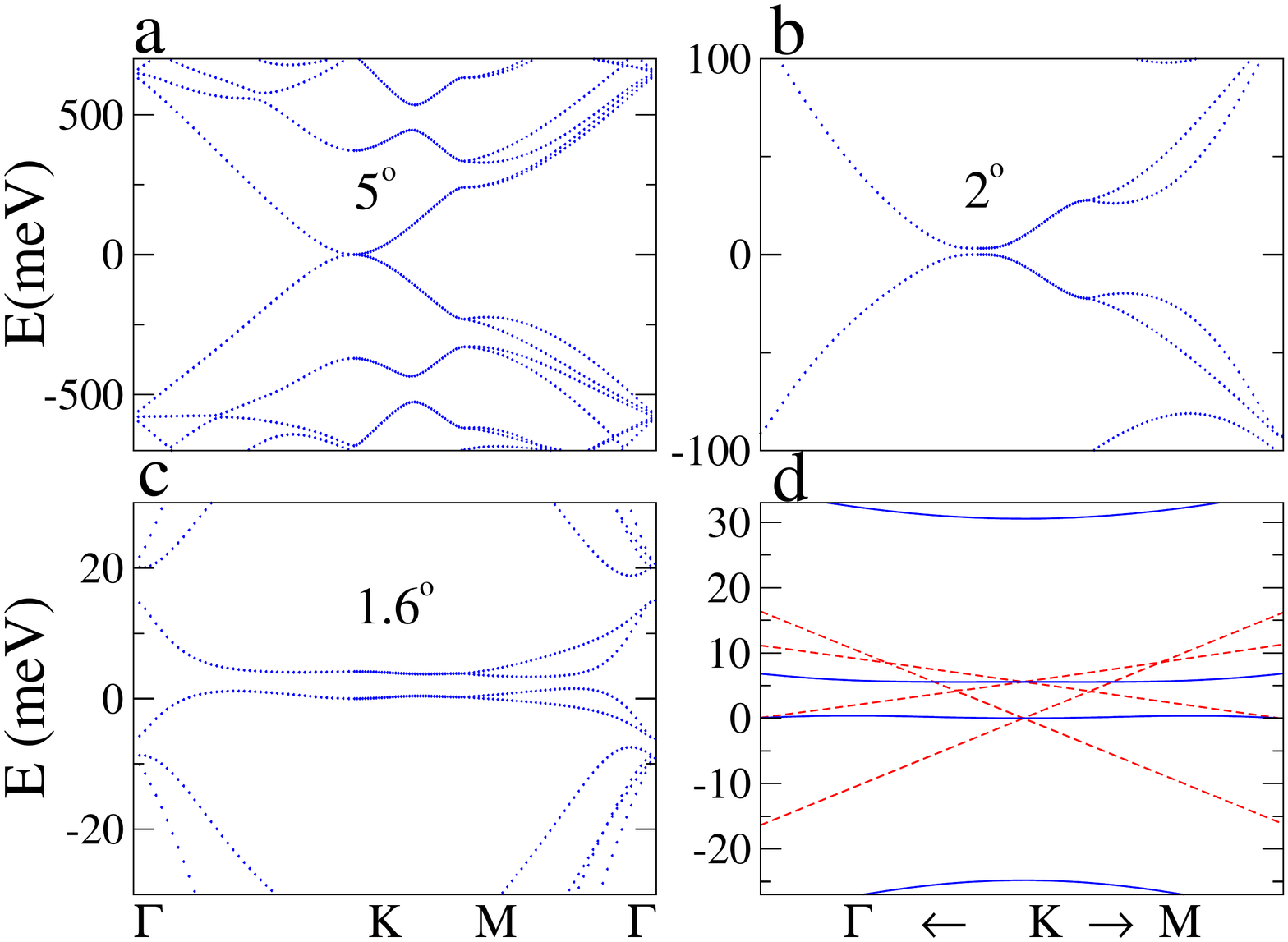}
\caption{\label{fig:TBS}
(a-c) Band structures calculated with a TB model including only one intralayer term ($\gamma_0$), one direct interlayer hopping ($\gamma_1$) between AB-stacked graphene and an exponential decay hopping between atoms in the inner twisted layers. There is no gap with this model for large angles. In d we show two calculations of the bands of DTBG with a rotation angle of 2.0$^{\rm o}$.  Dashed red lines:  the outer Bernal layers are uncoupled to the inner ones;  Continuous blue lines: outer Bernal-stacked layers coupled with a 1/10 fraction of $\gamma_1$. Notice that the size of the gap in the coupled case is exactly the shift of the Dirac cone of the twisted layers. 
}
   \end{figure}

In Fig.\ref{fig:TBS} (a-c) we show the results of the band structures for three angles with only nearest-neighbor intralayer hopping $\gamma_0$. The gap is very small for large angles, i.e., above 5$^{\rm o}$ but it becomes noticeable 
for angles below 2$^{\rm o}$. An almost perfect electron-hole symmetry can be observed, and the width of the flat bands obtained for the $1.16^{\rm o}$ is narrower at the K-M direction. Notice that in the full TB model the bands are narrower in the M-$\Gamma$ direction, as mentioned in Section \ref{res}.

In Fig. \ref{fig:TBS}d we explicitly show the origin of the gap in this model which is the second mechanism for gap opening explained in the main text. In dashed red lines we plot two coupled inner twisted layers and two uncoupled graphene outer layers. This system is like DTBG but with uncoupled outer layers. In this case we obtain two Dirac cones, one belonging to the twisted layers, with renormalized velocity, and the other being in fact two degenerate cones with the slope as in monolayer graphene. Notice the shift in energy of the twisted Dirac cone. 
We also superpose in blue lines a band structure calculation for the same system, but now for an small interlayer coupling of $0.1\gamma_1$  ($\gamma_1$ is the nearest interlayer hopping in AB-stacked bilayer graphene). The size of the gap in the parabolic bands is exactly the same as the shift of the twisted Dirac cone. The other two bands are the anti-bonding bands of AB bilayer graphene, now located near the Fermi level due to the smaller coupling. Importantly, the size of the gap does not change with  $\gamma_1$.


%
\end{document}